\documentclass[12pt,preprint]{aastex}
\newcommand{\beq}{\begin{equation}}
\newcommand{\eeq}{\end{equation}}
\newcommand{\beqa}{\begin{eqnarray}}
\newcommand{\eeqa}{\end{eqnarray}}

\shorttitle{Millisecond Pulsar Visibility}
\shortauthors{Venter \& De Jager}
\begin{document}

\title{EMPIRICAL CONSTRAINTS ON THE GR ELECTRIC FIELD ASSOCIATED WITH PSR~J0437-4715}
\author{C. VENTER\altaffilmark{1} AND O.C. DE JAGER\altaffilmark{2}}
\affil{Unit for Space Physics, North-West University, Potchefstroom Campus,\\ Private Bag X6001, Potchefstroom, 2520, South Africa}
\altaffiltext{1}{fskcv@puk.ac.za}
\altaffiltext{2}{fskocdj@puk.ac.za}

\begin{abstract}
We simulate the magnetosphere of the nearby millisecond pulsar PSR~J0437-4715, which is expected to have an unscreened electric potential due to the lack of magnetic pair production. We incorporate General Relativistic (GR) effects and study curvature radiation (CR) by primary electrons, but neglect inverse Compton (IC) scattering of thermal X-ray photons by these electrons. We find that the CR spectrum cuts off at energies below $\sim$ 17~GeV, well below the threshold of the H.E.S.S.\ telescope ($\lesssim$ 100~GeV), while other models predict a much higher cutoff of $\gtrsim$ 100~GeV. GR theory also predicts a relatively narrow pulse ($\beta^o \sim 0.2$ phase width) centered on the magnetic axis. EGRET observations above 100~MeV significantly constrain the application of the Muslimov \& Harding (1997) model for $\gamma$-ray production as a result of GR frame dragging, and ultimately its polar cap (PC) current and accelerating potential. Whereas the standard prediction of this pulsar's $\gamma$-ray luminosity due to GR frame dragging is $\sim$ 10\% of the spindown power, a non-detection by forthcoming H.E.S.S.\ observations will constrain it to $\lesssim$ 0.3\%, enforcing an even more severe revision of the accelerating electric field and PC current.
\end{abstract}

\section{INTRODUCTION}
Several authors have included General Relativistic (GR) frame dragging in models of pulsar 
magnetospheric structure and associated radiation and transport processes, 
recognising it to be a first order effect (see e.g.\ \citeauthor{MH97} \citeyear{MH97} (MH97); \citeauthor{drb01} \citeyear{drb01}).

\citet{u83} was the first to suggest that the low magnetic field strengths of millisecond pulsars (MSPs)
allow \mbox{$\gamma$-rays} up to at least 100~GeV to escape pair production. Most MSPs have (largely) unscreened 
electric fields due to the low optical depths of primary curvature \mbox{$\gamma$-rays}
for pair production in such low-B pulsar magnetospheres \citep[][HMZ02]{hmz02}. Radiation reaction limited curvature \mbox{$\gamma$-rays} up to about 100~GeV from MSPs
have been predicted \citep[HMZ02;][BRD00]{brd00}, making nearby MSPs such as 
PSR J0437-4715 \citep{j93} attractive targets for ground-based $\gamma$-ray groups 
\citep[BRD00;][]{ven04}. The unscreened case offers a  
test for fundamental GR electrodynamical derivations of the 
polar cap (PC) current and potential, without having to invoke additional modifications such
as pair formation fronts \citep[][HM98]{HM98} with associated slot gaps \citep{mh03} to explain 
most observations of canonical (high-B) pulsars.

The use of an unscreened GR electric field (see section \ref{sec:Efield}) for PSR~J0437-4715
(implied by its relatively low spindown power - HMZ02) was justified \textit{a posteriori} (see section \ref{sec:pairprod}). Several important parameters, most notably its mass and distance \citep{VS01}, are accurately known, making PSR J0437-4715 one of the closest pulsars to earth and probably much brighter and easier observable than other MSPs. Also, observations show that the radio and X-ray beams virtually coincide \citep{ZP02}, implying that the observer sweeps through the approximate center of the PC \citep{mj95,gk97}. 

In this paper, we investigate the effect of GR constraints on MSP spectral cutoffs, pulse profiles, integral flux and conversion efficiency of spindown power to \mbox{$\gamma$-ray} luminosity
by simulating (using a finite element approach) radiative and transport 
processes which occur in a pulsar magnetosphere. 

\section{THE UNSCREENED ELECTRIC FIELD AND RADIATIVE LOSSES}\label{sec:Efield}
We use the GR-corrected expressions for a static dipolar magnetic field (e.g.\ \citeauthor{MT92} \citeyear{MT92} (MT92); MH97) and curvature radius $\rho_c$ (e.g.\ HM98) of an oblique pulsar with magnetic moment $\mu = B_0R^3/2$ 
inclined at an angle $\chi$ relative to the spin axis. The value of the surface magnetic field (at the pole), $B_0$,
was solved for using (MH97)
\beq
\dot{E}_{\rm rot} \equiv I\Omega\dot{\Omega} \approx \frac{B_0^2\Omega^4 R^6}{6c^3f^2(1)},\label{eq:B}
\eeq
with $\dot{E}_{\rm rot}$ the spindown power, $\Omega$ the angular speed, $\dot{\Omega}$ the time-derivative thereof, $I$ the moment of
inertia, $R$ the stellar radius, $c$ the speed of light in vacuum and $f(\eta)$ defined by eq.~(25) of MT92.

The effect of GR frame dragging on the charge density, electric potential and hence the magnitude of $E_{||}$ (the electric field component parallel to the local magnetic field lines) was carefully modelled for the unscreened case,
since the optical depth for magnetic pair production above the PC is insignificantly small (see section~\ref{sec:pairprod}). The `near' and `distant' cases for $E_{||}$ (when $\eta \simeq 1$ and $\eta \gg 1$, with $\eta = r/R$), coincide at different points for different pulsar parameters. 
We use the same framework as Harding, Muslimov and Tsygan (MT92; MH97; HM98), with all the symbols
corresponding to their formalism. For the `near' case,
\beq
E_{||}^{\rm near} = -\frac{\Phi_0}{R}\left\{12\kappa\Theta_0^2s_1\cos\chi + 6s_2\Theta_0^3H(1)\delta(1)\sin\chi\cos\phi\right\},\label{eq:E_noUpper_Near}
\eeq
with the vacuum potential $\Phi_0 \equiv B_0\Omega R^2/c$, compactness parameter $\kappa = \epsilon I/MR^2$, 
$\epsilon = 2GM/Rc^2$, pulsar mass $M$, polar angle of last closed magnetic field line $\Theta(\eta) = [(\Omega R\eta/cf(\eta)]^{1/2}$ 
and $\Theta_0 \equiv \Theta(1)$ (HM98). 
Furthermore, in eq.~(\ref{eq:E_noUpper_Near}),
\beqa
s_1 & = & \sum_{i=1}^{\infty}\frac{J_0(k_i\xi)}{k_i^3J_1(k_i)}\left[1-e^{-\gamma_i(1)(\eta-1)}\right]\label{eq:e1}\\
s_2 & = & \sum_{i=1}^{\infty}\frac{J_1(\tilde{k}_i\xi)}{\tilde{k}_i^3J_2(\tilde{k}_i)}\left[1-e^{-\tilde{\gamma}_i(1)(\eta-1)}\right],\label{eq:e2}
\eeqa
with $k_i$ and $\tilde{k}_i$ the positive roots of the Bessel functions $J_0$ and $J_1$, $\xi \equiv \theta/\Theta(\eta)$ the normalised polar angle, $\phi$ the magnetic azimuthal angle, and $H(1)\delta (1) \approx 1$. (For $\gamma_i$ and $\tilde{\gamma}_i$, see eq.~[22] in HM98 and definitions following eq.~[43] in MT92). Note that $E_{||}^{\rm near}(\eta =1) = 0$ as required by the boundary conditions and that $E_{||}$ scales linearly with radial distance $\eta$ close to the stellar surface (derived from a Taylor expansion of eq.~[\ref{eq:e1}] and eq.~[\ref{eq:e2}] at $\eta \sim 1$). For the `far' case ($\eta > R_{PC}/R$), we use (HM98)
\beq
E_{||}^{\rm far}  \simeq  -\frac{\Phi_0}{R}\left(1-\xi^2\right)\Theta_0^2\left\{\frac{3}{2}\frac{\kappa}{\eta^4}\cos\chi+\frac{3}{8}\Theta(\eta)H(\eta)\delta(\eta)\xi\sin\chi\cos\phi\right\},\label{eq:Epar_noUpper}
\eeq
and for the corotating charge density $\rho_e$, we use eq.~(32) in MT92.

The change in the energy of a primary electron is given (when only the dominating curvature radiation (CR) component is considered, neglecting inverse Compton (IC) scattering and synchrotron radiation) by
\beq
\frac{dE}{dt} = e\beta_rcE_{||} - \frac{2}{3}\left(\frac{e^2c}{\rho_c^2}\right)\gamma^4,\label{eq:electron_power}
\eeq
with $e$ the electron charge, $\beta_r = v_e/c \sim 1$ the normalised electron speed and $\gamma$ the Lorentz factor. The photon energy $\epsilon_{\gamma}$ is set equal to the characteristic CR energy $\epsilon_c \equiv 1.5(\lambda_c/\rho_c)\gamma^3$ (in units of $m_ec^2$ - \citet{luo00}), with $\lambda_c = \hbar/m_ec \approx 3.86 \times 10^{-11}$ cm the Compton wavelength.

\section{PAIR PRODUCTION, SPECTRA AND CUTOFFS}\label{sec:pairprod}
According to HMZ02, the CR death line is at $\dot{E}_{\rm rot} \lesssim 10^{35}$ erg\,s$^{-1}$.
Evaluating $\dot{E}_{\rm rot} = -4\pi I\dot{P}/P^3 \sim 4 \times 10^{33}$ erg\,s$^{-1}$
(using the corrected intrinsic period derivative $\dot{P}$ - \citet{VS01}) suggests that no pair production will take place. Detailed modelling yields negligible optical depths, confirming this scenario. This is indeed fortunate because of the limited number of free parameters in this case. However, a low intensity of IC scattered UV photons/soft X-rays into the TeV range may contribute to a weak pair production component.

For the parameter ranges $R_6 \equiv R/10^6\;{\rm cm} = 1.3 - 1.7$ \citep[e.g.][]{KPR04}, $I_{45} \equiv I/10^{45}\;{\rm g.cm^2} = 1-3$ (e.g.\ HMZ02), and ($\chi$,$\zeta$) = (35\degr,40\degr) \citep{mj95}, ($\chi$,$\zeta$) = (20\degr,25\degr) \citep[e.g.][]{pz97} and ($\chi$,$\zeta$) = (20\degr,16\degr) \citep{gk97}, with $\zeta$ the observer angle, the maximum CR cutoff energy is obtained by using $R_6 = 1.3$, $I_{45} = 3$ and $\chi=20\degr$. We used $M=1.58M_\sun$ derived from Shapiro delays \citep{VS01}. This corresponds to $\kappa \sim 0.2$ and surface magnetic field strength $B_8 \equiv B_0/10^8\;{\rm G} \sim 7.2$ (see eq.~[\ref{eq:B}]). The relative altitude for maximum CR energy is obtained as $\eta \sim 1.47$ corresponding to a normalized field line colatitude of $\xi \sim 0.1$ and $\rho_c \sim 10^8$ cm, while the magnetic azimuth $\phi=0$ results in a maximum GR potential. The analytical expression for the maximum $\gamma$-ray energy is obtained by combining eq.~(\ref{eq:Epar_noUpper}), (\ref{eq:electron_power}), and the expression for $\epsilon_{\gamma}$, giving
\beq
\epsilon_{\gamma,\max} = \left(\frac{3}{2}\right)^{7/4}\lambda_c\left(\frac{\beta_r E_{||}}{e}\right)_{\rm max}^{3/4}\rho_c^{1/2} 
\lesssim 17\,\mathrm{GeV}.
\eeq

One of the most interesting predictions from MH97 is that the primary electron luminosity is given by (assuming $\chi \sim 0$):
\beq
L_{\rm prim,max}^{|\chi=0} \sim \frac{3}{4}\kappa(1-\kappa)\dot{E}_{\rm rot}.\label{eq:L_prim_max_old}
\eeq
It is important to note that the electric potential and charge density were derived assuming that electrons leave the PC with a speed equal to $c$. Even if the stellar injection speed $\beta_Rc \ll c$, it can be shown that the electrons will become relativistic very close to the neutron star surface, making maximum electron energies virtually independent of the injection speed (also: A.K.\ Harding 2004, personal communication). The bolometric particle luminosity of a single PC will therefore be given by (MH97)
\beq
L_{\rm prim} = \alpha c\int|\rho_e|\Phi\,dS,
\eeq
with $\Phi$ the electric potential and $dS$ the element of spherical surface cut by the last open field lines
at radial distance $r$. Integrating over $\xi$ and $\phi$, and letting $\eta \rightarrow \infty$, we obtain \citep{ven04}
\beq
L_{\rm prim,max} = L_{\rm prim,max}^{|\chi=0}\left\{\cos^2\chi + \left(\frac{3\Theta_0H(1)\left[\pi/2-\Theta_0H(1)\right]}{16\kappa(1-\kappa)}\right)\sin^2\chi\right\},\label{eq:L_prim_max_new}
\eeq
providing we adopt a value of $\Theta(\eta) = \pi/2$ for distances $\eta>c/\Omega R$. This result reduces back to eq.~(\ref{eq:L_prim_max_old}) when $\chi$ is set equal to zero. We calculated the maximum efficiency of conversion of pulsar spindown power into particle luminosity $L_{\rm prim,max}$ for $\chi=20\degr$ and $\chi=35\degr$, and obtained $\sim 2-11\%$ for PSR J0437-4715, for each PC (depending on $R$ and $I$, using $M = 1.58M_\sun$). 
We also obtained the bolometric photon luminosity $L_{\gamma}$ using a finite element (particle tracing) approach and integrating numerically over all photon energies and field lines from the surface to the light cylinder:
\beq
L_{\gamma} = \int_0^{\Theta_0}\!\!\!\!\int_0^{2\pi}\!\!\!\left(\dot{N}(\phi_R,\theta_R)\int_{r = R}^{r=c/\Omega}\!\!\!P_{\gamma}(\phi,\theta,r)\,dt\right)d\phi\,d\theta.
\label{eq:L_gamma}
\eeq
Here $P_{\gamma}$ is the CR photon power integrated over frequency and $\dot{N} = \rho_ec\,dS/e$ the number of particles ejected per second from a PC surface patch $dS$ at $r = R$ centered at $(\phi_R,\theta_R)$. 
Since we cannot start with $\beta_R=1$ (i.e.\ infinite Lorentz factor), we assumed values close to~1 and found convergent photon luminosities of $2-9\%$ of the spindown power (depending on $R$, $I$, $\chi$ and $\zeta$), i.e.\ $L_{\gamma}/L_{\rm prim,max} \sim 1$. This means that almost all particle luminosity is converted to photon luminosity as expected for strong radiation reaction. Radiation reaction, combined with further (weak) acceleration towards the light cylinder, result in a total residual electron power of $\sim 1-2.5\%$ of the spindown power at the light cylinder.

It should be noted that the fundamental unscreened expression for $E_{||}$ (eq.~[\ref{eq:Epar_noUpper}]) changes sign along $\sim 40\%$ of the magnetic field lines originating at the PC. This field reversal is most dominant when $\phi \sim \pi$, whereas no field field reversals occur for $\phi\sim 0$. Trapping of electrons may ensue at magnetic field lines along which the electric field reverses. We expect the system to reach a steady state as a result of the redistribution of charges along these field lines. These lines may become equipotential lines, or a reduced current may develop, resulting in the suppression of particle acceleration along them. This justifies our neglect of these field lines when calculating the pulse profiles, bolometric photon luminosity and integral flux.

\begin{figure}[ht]
    \plotone{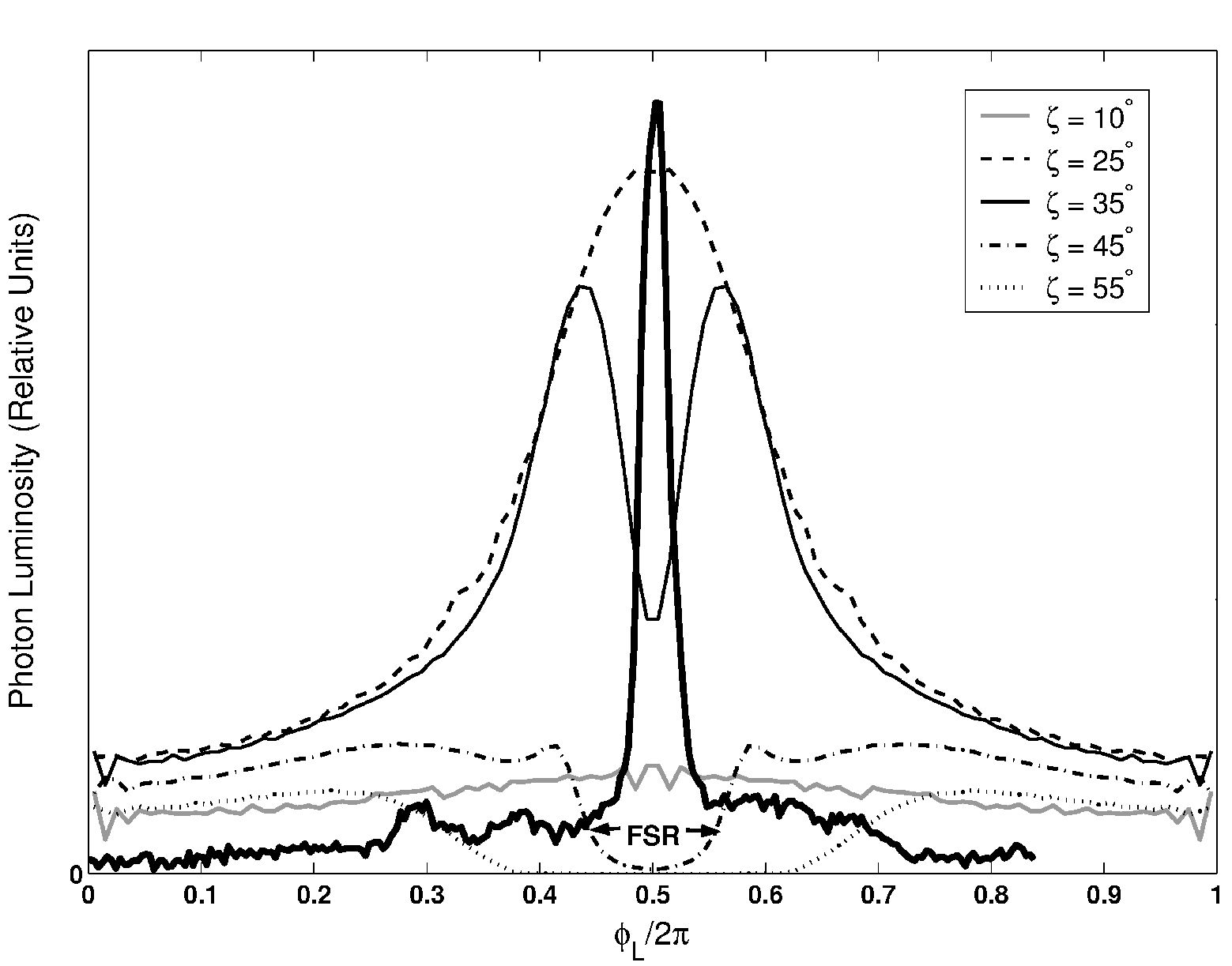}
    \caption{Photon luminosity (in relative units) vs.\
    observer pulse phase (with phase 0.5 corresponding either to $\phi = 0$ or $\phi = \pi$, depending on $\zeta$) 
    for PSR~J0437-4715 for different $\zeta$ (see legend). 
    The following parameters were assumed (see text for references): $P \approx 5.76$ ms 
    (period), $R_6 = 1.3$, $I_{45} = 1$, $M = 1.58\,M_{\sun}$ and $\chi = 35\degr$. The radio pulse at 4.6~GHz 
    (thick solid line - \citet{mj95}) is superimposed for reference (see www.atnf.csiro.au/research/pulsar/psrcat). 
    The ``valleys'' at observer phase $\sim $ 0.5 of the 
    lightcurves with $\zeta \geq \chi$ are probably due to electric field sign reversals (FSR), since the
    magnetic field lines where these reversals occur, were ignored (see text for details).
    \label{fig1}}
    \end{figure}

Figure \ref{fig1} shows the pulse profiles for different observer angles $\zeta$, for $\chi = 35\degr$. 
Maximum observed photon flux is obtained for $\zeta \sim \chi$ and for large values of $\cos\phi$ (as in eq.~[\ref{eq:Epar_noUpper}]). The `dip' in light curves with $\zeta \geq \chi$ near phase $\phi_L/2\pi \sim 0.5$ (where $\phi \sim \pi$) might be due to the sign reversal of the electric field, because the magnetic field lines where this sign reversal occur, were ignored as noted above.

The differential photon power $dL_{\gamma}(\phi_L,\zeta,E)/d\phi_Ld\zeta dE$ per phase bin $d\phi_L$, per observer angle bin $d\zeta$, per energy bin $dE$, is obtained by inserting the product of the ratios of indicator functions $I(\phi_L,\phi_L+d\phi_L)$, $I(\zeta,\zeta+d\zeta)$ and $I(E,E+dE)$ and their respective bin widths $d\phi_L, d\zeta$ and $dE$ in the integrand of eq.~(\ref{eq:L_gamma}). This allows us to compare the expected integral photon flux with EGRET upper limits above 100 MeV and 1 GeV \citep{f95}, as well as with forthcoming H.E.S.S.\ observations of this pulsar \citep{ven04}. Note that the imaging threshold energy of H.E.S.S.\ is $\sim$~100~GeV \citep{h01}, although a non-imaging ``pulsar trigger'' for timing studies down to $\gtrsim$~50~GeV can be employed for pulsar studies with H.E.S.S.\ \citep{d01}.

The phase-averaged photon flux (as would be seen on a DC skymap) for a single PC may be calculated by 
\beq
\overline{F}^o_{\gamma}(>\!\!E) = \frac{\beta^o}{d^2\overline{\Delta\Omega}^o} \int_{E}^{\infty}\int_\zeta^{\zeta+d\zeta}\int_0^{2\pi}\frac{1}{E^\prime}\left[\frac{dL(\phi_L^\prime,\zeta^\prime,E^\prime)}{d\phi_L^\prime d\zeta^\prime dE^\prime}\right]d\phi_L^\prime d\zeta^\prime dE^\prime,
\eeq 
with distance $d=139$ pc, $\beta^o = \Delta\phi_L/2\pi$, $\Delta\phi_L$ the pulse width in radians, 
$\overline{\Delta\Omega}^o(>\!\!E) = \sin\zeta d\zeta \Delta\phi_L$ the beaming solid angle, and $d\zeta$ taken arbitrarily small. Only one PC is expected to be seen, given the relative orientations of the magnetic axis and observer line-of-sight to the spin axis. The superscript `o' will be used to indicate quantities applicable to an observer with $\zeta \in (\zeta,\zeta+d\zeta)$.

\begin{figure}[ht]
    \plotone{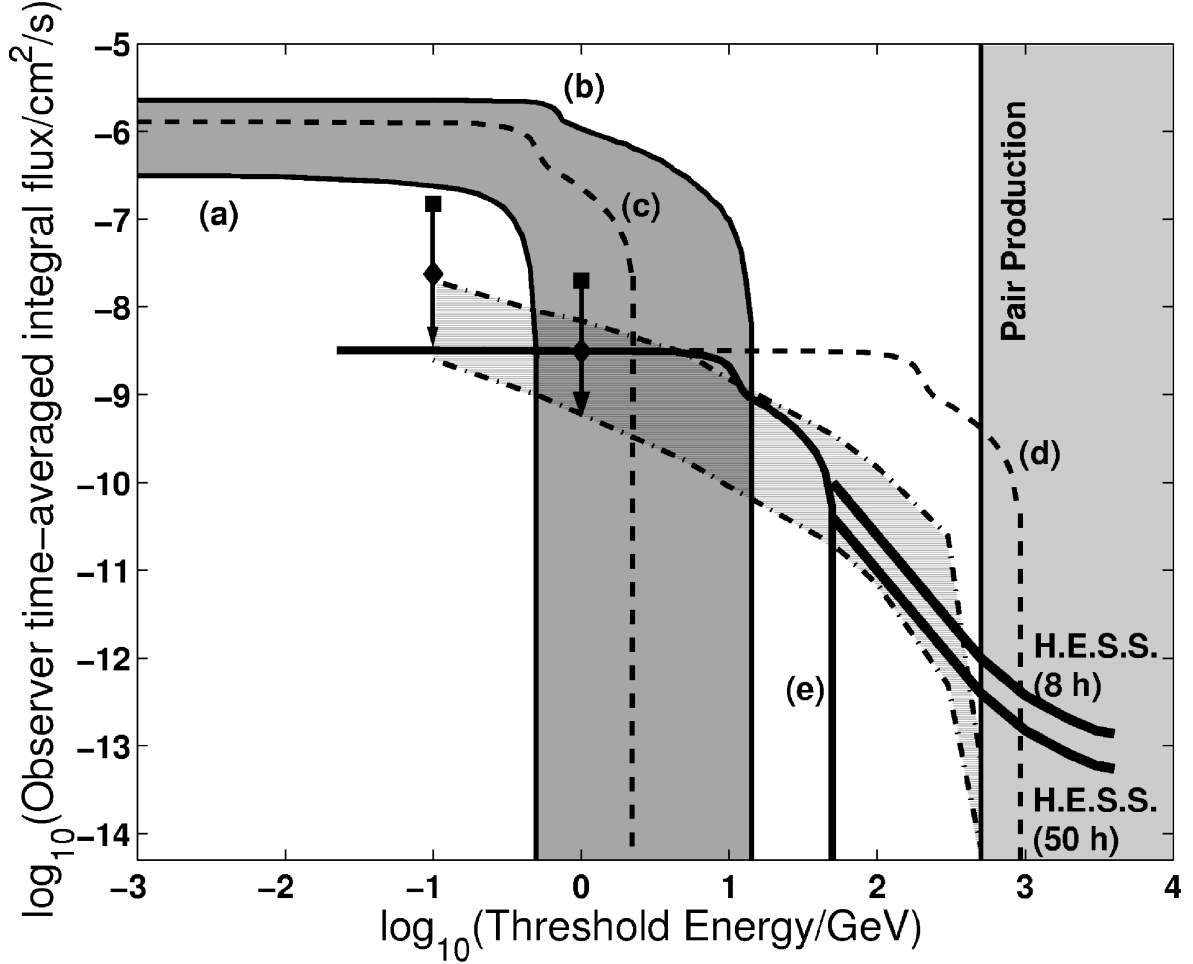}
    \caption{Observer time-averaged integral flux vs.\ threshold energy.
    Curve~(a), for which $R_6  = 1.7$, $I_{45} = 1$, $\chi = 35\degr$ and $\zeta = 40\degr$ and curve~(b) for which
    $R_6 = 1.3$, $I_{45} = 3$, $\chi = 20\degr$ and $\zeta = 16\degr$, define
    a ``confidence band'' wherein the integral flux is expected to lie according to the GR model discussed
    in this paper. Curve~(c), for which $R_6 = 1.5$, $I_{45} = 2$, $\chi = 20\degr$, and $\zeta = 16\degr$, 
    represents an intermediate curve. Curve~(d) is curve~(c) scaled with scale factor $\lambda = 400$, while 
    curve~(e) is curve~(d) shifted to the left (see text for details). The band with dot-dashed curves is that of
    BRD00 for PSR J0437-4715 for their `Model A'. The squares represent EGRET integral flux
    upper limits \citep{f95}, while the diamonds represent these upper limits reduced by a factor $\sqrt{5}$,
    appropriate for a beam with main pulse width of $\sim 0.2$. Also indicated are the H.E.S.S.\ sensitivities
    for 50 hours \citep{h04} and 8 hours observation time, and the energy above which pair production is
    expected to take place (BRD00).
   \label{fig2}}
    \end{figure}

The energy spectrum $dL/dE$ due to CR is quite hard, resulting in a constant time-averaged integrated photon flux
$\overline{F}^o_{\gamma}(>\!\!E)$, seen by the observer, as shown in figure~\ref{fig2} (e.g.\ curves~(a) and (b)). The 100~MeV and 1~GeV EGRET flux upper limits from \citet{f95} are indicated by the squares on figure~\ref{fig2}, which clearly constrain the flux band defined by (a) and (b). If we define an {\it a priori} phase interval of $\beta^o \sim 0.2$, centered on the radio pulse, and recalculate the EGRET flux upper limits from the factor five ($=1/\beta^o$) reduced skymap background, we should get the even more constraining upper limits given by the diamonds in figure~\ref{fig2}. We therefore have to revise the predicted fluxes for PSR~J0437-4715 and we do so based on the following scaling argument: 
If we assume that the particle and hence $\gamma$-ray luminosity only scales with the spindown power and neutron star compactness, as in eq.~(\ref{eq:L_prim_max_old}) and  eq.~(\ref{eq:L_prim_max_new}), i.e. the product of the current and voltage for such a pair-starved pulsar is a constant as predicted by eq.~(\ref{eq:L_prim_max_old}), we may scale the set of curves~(a) through (c) (according to this condition of a constant photon luminosity $L^o_{\gamma}$) in terms of the limiting voltage and hence the cutoff energy to give  $\overline{F}^o_{\gamma,1}(>\!\!E_1)\times E_{\rm cutoff,1} \sim \overline{F}^o_{\gamma,2}(>\!\!E_2)\times E_{\rm cutoff,2}$ (for constant $\beta^o$ and $\overline{\Delta\Omega}^o$, and energies $E_1 < E_2$; $E_{\rm cutoff,1} < E_{\rm cutoff,2}$). In particular, when curve~(c) is scaled according to $E_{\rm cutoff,2} = \lambda E_{\rm cutoff,1}$, implying $\overline{F}^o_{\gamma,2}(>\!\!E_2) \sim \overline{F}^o_{\gamma,1}(>\!\!E_1)/\lambda$, with $\lambda = 400$, curve~(d) is obtained, which no longer violates the revised EGRET upper limit at 1~GeV, but the cutoff energy then shifts up to $\sim$ 1~TeV. Furthermore, if curve~(d) is now translated so that the energy cutoff also falls below the H.E.S.S.\ sensitivity curves, curve~(e) is obtained, which would be consistent with both EGRET and H.E.S.S. (if the latter instrument does not detect this pulsar). Also shown is the flux band calculated for PSR J0437-4715 by BRD00. Again, for a power-law photon spectrum with exponential cutoff, it can be shown that $\overline{F}^o_{\gamma}(>\!\!E_1) E_{\rm cutoff} \sim \beta^o L^o_{\gamma}/d^2\overline{\Delta\Omega}^o$ (assuming $E_1 \ll E_{\rm cutoff}$, and $\overline{F}^o_{\gamma}(>\!\!E)$ having a flat slope due to CR). In order to constrain PSR~J0437-4715's bolometric photon luminosity by forthcoming H.E.S.S.\ observations, we postulate that $L_{\gamma} = \alpha L^o_{\gamma}$, where $\alpha = \alpha(\chi,\zeta)\gg 1$ is a geometrical factor
correcting from the incremental luminosity corresponding to the observer's line-of-sight, to the total $\gamma$-ray luminosity of the pulsar. It then follows that $L_{\gamma} \sim x\overline{F}^o_{\gamma}(>\!\!E_1) E_{\rm cutoff}$, with $x(\chi,\zeta) = \alpha d^2 2\pi \sin\zeta d\zeta$, which was found to be more or less constant for the same $\chi$ and $\zeta$. A non-detection by H.E.S.S., as implied by curve~(e), leads to a $\gamma$-ray luminosity of $\lesssim 0.003\dot{E}_{\rm rot}$. This value should be compared with the prediction of $L_{\gamma} \sim 3\times 10^{-5}\dot{E}_{\rm rot}$ given by \citet{RD99} for a canonical pulsar with $P = 1$~ms and $B_0 = 10^9$~G and with
$L_{\gamma} \sim 0.04\dot{E}_{\rm rot}$ predicted for pair-starved pulsars with off-beam geometry (using $P \approx 5.76$ ms and $\dot{E}_{\rm rot} \sim 4\times10^{33}$ erg.s$^{-1}$ - \citet{mh04}).

\section{CONCLUSIONS}
CR cutoff energies for MSPs such as PSR J0437-4715 were predicted to be in the range $50-100$~GeV 
by HMZ02 and BRD00, making proposals for ground-based telescopes with imaging thresholds near 100~GeV
(e.g. H.E.S.S.\ \citep{h01} and CANGAROO \citep{yyy02}) attractive. 
From the present GR theory, one would conclude that these telescopes may not be able to see the
spectral tail corresponding to the intense primary CR component, since the hard primary CR spectrum does not
extend to energies above $\sim$~20~GeV, as verified by both analytical and numerical (finite element) approaches. 
An IC component resulting from TeV electrons scattering the UV/soft X-rays from the surface of 
PSR J0437-4715 may however still be detectable, although this prediction by BRD00 should also be re-evaluated within a GR electrodynamical framework. 
However, it is quite obvious that the predicted time-averaged observer flux violates the EGRET upper limit at 100~MeV, impyling a revision of the existing theory. Forthcoming H.E.S.S.\ and future GLAST observations will help to constrain the $\gamma$-ray luminosity, and therefore the accelerating electric field.

\acknowledgments

The authors would like to acknowledge useful discussions with A.K.~Harding, B.~Rudak and A.~Konopelko. This publication is based upon work supported by the South African National Research Foundation under
Grant number 2053475.

\clearpage

\end{document}